\documentclass[12pt,fleqn]{book} % Default font size and left-justified equations
\usepackage{tcolorbox}
\usepackage{hyperref}
\usepackage{url}
\usepackage[utf8]{inputenc}
\usepackage[top=1.0in,bottom=1.0in,left=1.0in,right=1.0in,headsep=10pt,letterpaper]{geometry} % Page margins
\usepackage{xcolor} % Required for specifying colors by name
\definecolor{ocre}{RGB}{52,177,201} % Define the orange color used for highlighting throughout the book

\usepackage{avant} % Use the Avantgarde font for headings
\usepackage{mathptmx} % Use the Adobe Times Roman as the default text font together with math symbols from the Sym­bol, Chancery and Computer Modern fonts

\usepackage{microtype} % Slightly tweak font spacing for aesthetics
\usepackage[utf8]{inputenc} % Required for including letters with accents
\usepackage[T1]{fontenc} % Use 8-bit encoding that has 256 glyphs

\usepackage{titlesec} % Allows customization of titles

\usepackage{graphicx} % Required for including pictures
\graphicspath{{Pictures/}{figures/cosmo/}} % Specifies the directory where pictures are stored

\usepackage{lipsum} % Inserts dummy text

\usepackage{tikz} % Required for drawing custom shapes

\usepackage[english]{babel} % English language/hyphenation

\usepackage{enumitem} % Customize lists
\setlist{nolistsep} % Reduce spacing between bullet points and numbered lists

\usepackage{booktabs} % Required for nicer horizontal rules in tables

\usepackage{eso-pic} % Required for specifying an image background in the title page

\usepackage{titletoc} % Required for manipulating the table of contents

\contentsmargin{0cm} % Removes the default margin
\titlecontents{chapter}[1.25cm] % Indentation
{\addvspace{15pt}\large\sffamily\bfseries} % Spacing and font options for chapters
{\color{ocre!60}\contentslabel[\Large\thecontentslabel]{1.25cm}\color{ocre}} % Chapter number
{}
{\color{ocre!60}\normalsize\sffamily\bfseries\;\titlerule*[.5pc]{.}\;\thecontentspage} % Page number
\titlecontents{section}[1.25cm] % Indentation
{\addvspace{5pt}\sffamily\bfseries} % Spacing and font options for sections
{\contentslabel[\thecontentslabel]{1.25cm}} % Section number
{}
{\sffamily\hfill\color{black}\thecontentspage} % Page number
[]
\titlecontents{subsection}[1.25cm] % Indentation
{\addvspace{1pt}\sffamily\small} % Spacing and font options for subsections
{\contentslabel[\thecontentslabel]{1.25cm}} % Subsection number
{}
{\sffamily\;\titlerule*[.5pc]{.}\;\thecontentspage} % Page number
[]

\titlecontents{lsection}[0em] % Indendating
{\footnotesize\sffamily} % Font settings
{}
{}
{}

\titlecontents{lsubsection}[.5em] % Indentation
{\normalfont\footnotesize\sffamily} % Font settings
{}
{}
{}

\usepackage{fancyhdr} % Required for header and footer configuration

\fancypagestyle{plain}{\fancyhead{}} % Style for when a plain pagestyle is specified

\makeatletter
\renewcommand{\cleardoublepage}{
\clearpage\ifodd\c@page\else
\hbox{}
\vspace*{\fill}
\thispagestyle{empty}
\newpage
\fi}

\usepackage{amsmath,amsfonts,amssymb,amsthm} % For math equations, theorems, symbols, etc

\newtheoremstyle{ocrenumbox}% % Theorem style name
{0pt}% Space above
{0pt}% Space below
{\normalfont}% % Body font
{}% Indent amount
{\small\bf\sffamily\color{ocre}}% % Theorem head font
{\;}% Punctuation after theorem head
{0.25em}% Space after theorem head
{\small\sffamily\color{ocre}\thmname{#1}\nobreakspace\thmnumber{\@ifnotempty{#1}{}\@upn{#2}}% Theorem text (e.g. Theorem 2.1)
\thmnote{\nobreakspace\the\thm@notefont\sffamily\bfseries\color{black}---\nobreakspace#3.}} % Optional theorem note
\newtheoremstyle{blacknumex}% Theorem style name
{5pt}% Space above
{5pt}% Space below
{\normalfont}% Body font
{} % Indent amount
{\small\bf\sffamily}% Theorem head font
{\;}% Punctuation after theorem head
{0.25em}% Space after theorem head
{\small\sffamily{\tiny\ensuremath{\blacksquare}}\nobreakspace\thmname{#1}\nobreakspace\thmnumber{\@ifnotempty{#1}{}\@upn{#2}}% Theorem text (e.g. Theorem 2.1)
\thmnote{\nobreakspace\the\thm@notefont\sffamily\bfseries---\nobreakspace#3.}}% Optional theorem note

\newtheoremstyle{blacknumbox} % Theorem style name
{0pt}% Space above
{0pt}% Space below
{\normalfont}% Body font
{}% Indent amount
{\small\bf\sffamily}% Theorem head font
{\;}% Punctuation after theorem head
{0.25em}% Space after theorem head
{\small\sffamily\thmname{#1}\nobreakspace\thmnumber{\@ifnotempty{#1}{}\@upn{#2}}% Theorem text (e.g. Theorem 2.1)
\thmnote{\nobreakspace\the\thm@notefont\sffamily\bfseries---\nobreakspace#3.}}% Optional theorem note

\newtheoremstyle{ocrenum}% % Theorem style name
{5pt}% Space above
{5pt}% Space below
{\normalfont}% % Body font
{}% Indent amount
{\small\bf\sffamily\color{ocre}}% % Theorem head font
{\;}% Punctuation after theorem head
{0.25em}% Space after theorem head
{\small\sffamily\color{ocre}\thmname{#1}\nobreakspace\thmnumber{\@ifnotempty{#1}{}\@upn{#2}}% Theorem text (e.g. Theorem 2.1)
\thmnote{\nobreakspace\the\thm@notefont\sffamily\bfseries\color{black}---\nobreakspace#3.}} % Optional theorem note
\makeatother

\newcounter{dummy}
\numberwithin{dummy}{section}
\theoremstyle{ocrenumbox}
\newtheorem{theoremeT}[dummy]{Theorem}

\newtheorem{exerciseT}{Exercise}[chapter]
\theoremstyle{blacknumex}
\newtheorem{exampleT}{Example}[chapter]
\theoremstyle{blacknumbox}

\newtheorem{definitionT}{Definition}[section]
\newtheorem{corollaryT}[dummy]{Corollary}
\theoremstyle{ocrenum}

\RequirePackage[framemethod=default]{mdframed} % Required for creating the theorem, definition, exercise and corollary boxes

\newmdenv[skipabove=7pt,
skipbelow=7pt,
backgroundcolor=black!5,
linecolor=ocre,
innerleftmargin=5pt,
innerrightmargin=5pt,
innertopmargin=5pt,
leftmargin=0cm,
rightmargin=0cm,
innerbottommargin=5pt]{tBox}

\newmdenv[skipabove=7pt,
skipbelow=7pt,
rightline=false,
leftline=true,
topline=false,
bottomline=false,
backgroundcolor=ocre!10,
linecolor=ocre,
innerleftmargin=5pt,
innerrightmargin=5pt,
innertopmargin=5pt,
innerbottommargin=5pt,
leftmargin=0cm,
rightmargin=0cm,
linewidth=4pt]{eBox}	

\newmdenv[skipabove=7pt,
skipbelow=7pt,
rightline=false,
leftline=true,
topline=false,
bottomline=false,
linecolor=ocre,
innerleftmargin=5pt,
innerrightmargin=5pt,
innertopmargin=0pt,
leftmargin=0cm,
rightmargin=0cm,
linewidth=4pt,
innerbottommargin=0pt]{dBox}	

\newmdenv[skipabove=7pt,
skipbelow=7pt,
rightline=false,
leftline=true,
topline=false,
bottomline=false,
linecolor=gray,
backgroundcolor=black!5,
innerleftmargin=5pt,
innerrightmargin=5pt,
innertopmargin=5pt,
leftmargin=0cm,
rightmargin=0cm,
linewidth=4pt,
innerbottommargin=5pt]{cBox}

\makeatletter
\renewcommand{\@seccntformat}[1]{\llap{\textcolor{ocre}{\csname the#1\endcsname}\hspace{1em}}}
\renewcommand{\section}{\@startsection{section}{1}{\z@}
{-2ex \@plus -1ex \@minus -.2ex}
{1ex \@plus.1ex }
{\normalfont\large\sffamily\bfseries}}
\renewcommand{\subsection}{\@startsection {subsection}{2}{\z@}
{-2ex \@plus -0.1ex \@minus -.2ex}
{0.5ex \@plus.2ex }
{\normalfont\sffamily\bfseries}}
\renewcommand{\subsubsection}{\@startsection {subsubsection}{3}{\z@}
{-2ex \@plus -0.1ex \@minus -.2ex}
{.2ex \@plus.2ex }
{\normalfont\small\sffamily\bfseries}}
\renewcommand\paragraph{\@startsection{paragraph}{4}{\z@}
{-2ex \@plus-.2ex \@minus .2ex}
{.1ex}
{\normalfont\small\sffamily\bfseries}}

\usepackage{hyperref}
\hypersetup{hidelinks,backref=true,pagebackref=true,hyperindex=true,colorlinks=false,breaklinks=true,urlcolor= ocre,bookmarks=true,bookmarksopen=false,pdftitle={Title},pdfauthor={Author}}

\newcommand{\thechapterimage}{}
\newcommand{\chapterimage}[1]{\renewcommand{\thechapterimage}{#1}}

\def\thechapter{\arabic{chapter}}
\def\@makechapterhead#1{
\thispagestyle{empty}
{\centering \normalfont\sffamily
\ifnum \c@secnumdepth >\m@ne
\if@mainmatter
\startcontents
\begin{tikzpicture}[remember picture,overlay]
\node at (current page.north west)
{\begin{tikzpicture}[remember picture,overlay]
\node[anchor=north west,inner sep=0pt] at (0,0) {\includegraphics[width=\paperwidth]{\thechapterimage}};
\draw[anchor=west] (5cm,-9cm) node [rounded corners=20pt,fill=ocre!10!white,text opacity=1,draw=ocre,draw opacity=1,line width=1.5pt,fill opacity=.6,inner sep=12pt]{\huge\sffamily\bfseries\textcolor{black}{\thechapter. #1\strut\makebox[22cm]{}}};
\end{tikzpicture}};
\end{tikzpicture}}
\par\vspace*{230\p@}
\fi
\fi}

\def\@makeschapterhead#1{
\thispagestyle{empty}
{\centering \normalfont\sffamily
\ifnum \c@secnumdepth >\m@ne
\if@mainmatter
\begin{tikzpicture}[remember picture,overlay]
\node at (current page.north west)
{\begin{tikzpicture}[remember picture,overlay]
\node[anchor=north west,inner sep=0pt] at (0,0) {\includegraphics[width=\paperwidth]{\thechapterimage}};
\draw[anchor=west] (5cm,-6cm) node [rounded corners=20pt,fill=ocre!10!white,fill opacity=.6,inner sep=12pt,text opacity=1,draw=ocre,draw opacity=1,line width=1.5pt]{\LARGE\sffamily\bfseries\textcolor{black}{#1\strut\makebox[22cm]{}}};
\end{tikzpicture}};
\end{tikzpicture}}
\par\vspace*{130\p@}
\fi
\fi
}
\makeatother

\usepackage{graphicx}
\graphicspath{{./figures/}}
\usepackage{appendix}
\usepackage{latexsym,amsmath,amsfonts,amssymb,booktabs}
\usepackage[font=small]{caption}
\usepackage{slashed,upgreek,amscd,cancel,tensor,color}
\usepackage{adjustbox}
\usepackage[numbers,sort&compress,square]{natbib}
\usepackage{epsfig,latexsym}
\usepackage{url}
\numberwithin{equation}{section}% numera le equazioni seconde le sezioni , e.g. 1.15 invece che consecutivamente; anche le appendici, eq. (A.1) etc. Richiede amsmath
\usepackage{doi}
\usepackage{subcaption}
\usepackage{mathtools}
\usepackage{upgreek}
\usepackage{stfloats}
\usepackage{afterpage}
\usepackage{multirow}

\begin{document}

%-------- coverpage, copyright and table of contents

%\chapterimage{intro.jpg} % Chapter heading image
\chapterimage{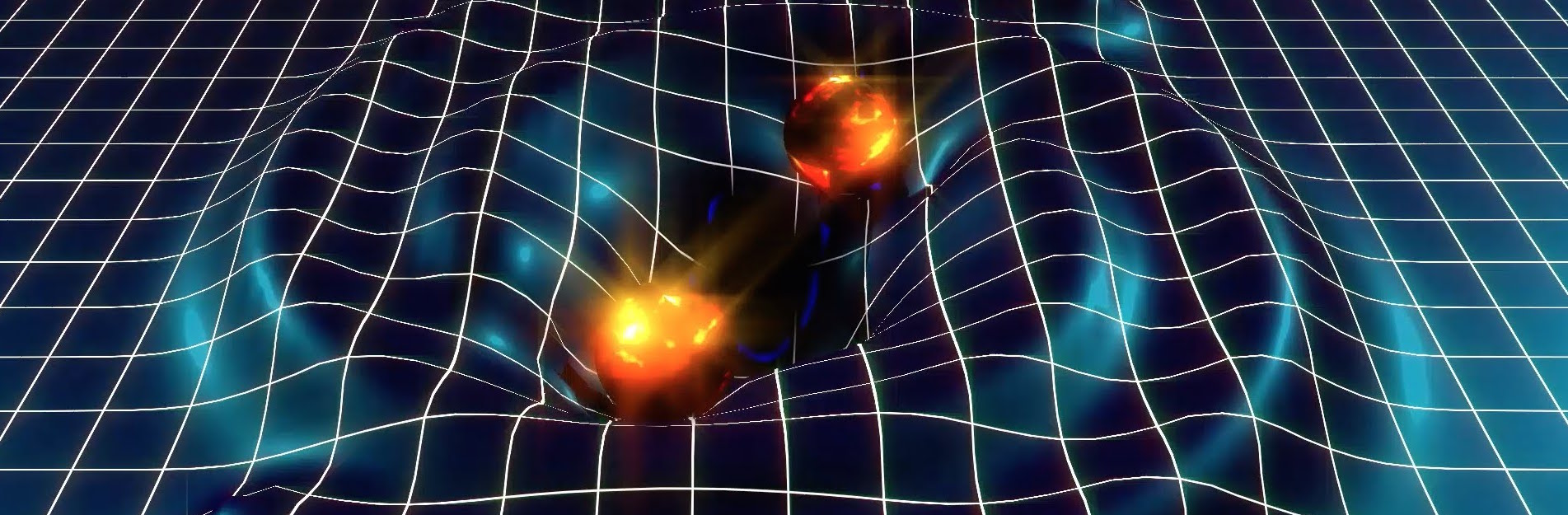} % Chapter heading image
\chapter*{\Large Extreme Gravity and Fundamental Physics}

\begin{center}
\Large
\textbf{Astro2020 Science White Paper} \linebreak

\vskip-0.5cm
%\vspace{-2cm} 

EXTREME GRAVITY AND FUNDAMENTAL PHYSICS \linebreak

\end{center} 

\normalsize

\vskip-0.5cm

\noindent \textbf{Thematic Areas:} 
\begin{itemize}
\item Cosmology and Fundamental Physics 
\item Multi-Messenger Astronomy and Astrophysics
\end{itemize}

% Planetary Systems 
% Star and Planet Formation 
% Formation and Evolution of Compact Objects 
% Cosmology and Fundamental Physics 
% Stars and Stellar Evolution 
% Resolved Stellar Populations and their Environments 
% Galaxy Evolution   
% Multi-Messenger Astronomy and Astrophysics 

\vspace{0.50cm}

\noindent \textbf{Principal Author}: 
\newline
\vspace{-0.5cm}

 Name:	B.S. Sathyaprakash
 \newline
\vspace{-0.5cm}

Institution: The Pennsylvania State University
 \newline
\vspace{-0.5cm}

Email: bss25@psu.edu
 \newline
\vspace{-0.5cm}

Phone: +1-814-865-3062 
 \newline
  
\noindent 
\textbf{Lead Co-authors:} 
Alessandra Buonanno (Max Planck Institute for Gravitational Physics, Potsdam and University of Maryland), Luis Lehner (Perimeter Institute), Chris Van Den Broeck (NIKHEF)\\

\noindent 
P. Ajith (International Centre for Theoretical Sciences), 
Archisman Ghosh (NIKHEF), 
Katerina Chatziioannou (Flatiron Institute), 
Paolo Pani (Sapienza University of Rome), 
Michael P\"urrer (Max Planck Institute for Gravitational Physics, Potsdam), 
Sanjay Reddy (Institute for Nuclear Theory, University of Washington, Seattle),
Thomas P. Sotiriou (The University of Nottingham), 
Salvatore Vitale (MIT), 
Nicolas Yunes (Montana State University), 
K.G. Arun (Chennai Mathematical Institute), 
Enrico Barausse (Institut d'Astrophysique de Paris), 
Masha Baryakhtar (Perimeter Institute),
Richard Brito (Max Planck Institute for Gravitational Physics, Potsdam), 
Andrea Maselli (Sapienza University of Rome),
Tim Dietrich (NIKHEF), 
William East (Perimeter Institute), 
Ian Harry (Max Planck Institute for Gravitational Physics, Potsdam and University of Portsmouth),
Tanja Hinderer (University of Amsterdam), 
Geraint Pratten (University of Balearic Islands and University of Birmingham), 
Lijing Shao (Kavli Institute for Astronomy and Astrophysics, Peking University),
Maaretn van de Meent (Max Planck Institute for Gravitational Physics, Potsdam), 
Vijay Varma (Caltech),
Justin Vines (Max Planck Institute for Gravitational Physics, Potsdam),  
Huan Yang (Perimeter Institute and U Guelph),
Miguel Zumalacarregui (U California, Berkely, and IPhT, Saclay)\\

\noindent 
Click here for \href{https://docs.google.com/spreadsheets/d/1uNKEW77Fm-_nc21_3jSOX-P4ecRxsxA5UgDOD4bkG9Y/edit#gid=112581173}{\bf other co-authors and supports}
%(names and institutions)

 % \linebreak

%\textbf{Abstract :}

%ADD ABSTRACT
\clearpage

\section*{Extreme Gravity and Fundamental Physics}
In general relativity, gravitational waves are non-stationary solutions of Einstein's equations arising as a result of time-varying quadrupole and higher-order multipole moments that translate into freely propagating oscillations in the fabric of spacetime \cite{Einstein:1918}. They emanate from regions of strong gravity and relativistic motion, yet the waves carry uncorrupted signature of their sources. They interact very weakly with matter and are hardly dispersed as they propagate from their sources to Earth, making them ideal for studying the dynamics of spacetime geometry \cite{thorne.k:1987,Sathyaprakash:2009xs}.

On September 14, 2015 the twin LIGO instruments at Hanford and Livingston made the first {\em direct} detection of gravitational waves \cite{Abbott:2016blz}.  Dubbed {\sc GW150914}, the waves were observed for 200 milliseconds and came from the final stages of the inspiral and merger of a binary black hole system at a distance of $\sim 450$ Mpc. To date LIGO and Virgo in Italy have detected ten binary black hole mergers \cite{LIGOScientific:2018mvr} that have helped to probe strong field gravity at unprecedented levels.

On August 17, 2017 LIGO and Virgo made another monumental discovery, this time the inspiral and coalescence of a pair of neutron stars \cite{TheLIGOScientific:2017qsa}.  Fermi Gamma-ray Space Telescope and the International Gamma-Ray Astrophysics Laboratory, both observed short gamma ray bursts $1.7$ seconds after LIGO's discovery \cite{GBM:2017lvd}, thus confirming the long-held conjecture that merging binary neutron stars are progenitors of short gamma ray bursts.

\begin{tcolorbox}[standard jigsaw,colframe=ocre,colback=ocre!10!white,opacityback=0.6,coltext=black]
Future gravitational-wave observations will enable unprecedented and unique science in {\it extreme gravity and fundamental physics,} that form the core topics of the Thematic Area 7 of Astro-2020 decadal survey.
\vskip5pt
\begin{itemize}[leftmargin=*]
\item {\bf The nature of gravity.} Can we prove Einstein wrong? What building-block principles and symmetries in nature invoked in the description of gravity can be challenged?
\item {\bf The nature of dark matter.} Is dark matter composed of particles, dark objects or modifications of gravitational interactions?
\item {\bf The nature of compact objects.} Are black holes and neutron stars the only astrophysical extreme compact objects in the Universe? What is the equation of state of densest matter?
\end{itemize}
\end{tcolorbox}

These detections have ushered in a new era of fundamental physics.  Gravitational-wave (GW) observations can be used for understanding not just the sky but also in testing general relativity in dynamical spacetimes ~\cite{Yunes:2016jcc,Berti:2018vdi,TheLIGOScientific:2016src,Abbott:2018lct} and in providing insights into the nature of matter under extreme physical conditions of gravity, density, and pressure~\cite{De:2018uhw, Tews:2018chv, Annala:2017llu,Abbott:2018exr}.  Advanced LIGO and Virgo will only be first steps in this new endeavor that is guaranteed to change our perception of the Universe in the coming decades.  Indeed, the next generation of GW observatories, such as the Einstein Telescope and Cosmic Explorer (referred to as 3G), will witness merging black holes and neutron stars when the Universe was still in its infancy assembling its first stars and black holes.  At such sensitivity levels we can expect to measure extremely bright events that could reveal subtle signatures of new physics. 3G observatories promise to deliver data that could transform the landscape of physics, addressing some of the most pressing problems in fundamental physics and strong gravitational fields.

\section*{The nature of gravity.}
Probing the nature of gravity and its possible implications on fundamental physics is a high-reward, even if uncertain, prospect of gravitational-wave observations.
To our knowledge, astrophysical black holes and relativistic stars exhibit the largest curvature of spacetime accessible to us. They are, therefore, ideal systems to observe the behavior of spacetimes under the most extreme gravitational conditions.
New physics indicative of departures from the basic tenants of General Relativity (GR) could reveal itself in high fidelity waveforms expected to be observed
in the next generation of detectors.

%\noindent
%\begin{minipage}{0.30\textwidth}
Such signals would provide a unique access to extremely warped spacetimes and gain invaluable insights on GR or what might replace it as the theory of gravity governing such systems. The adjacent diagram provides a perspective of the reach of different missions/facilities and their target regime with respect to characteristic spacetime curvature ($R$) and gravitational potential $\Phi$ (which for binary systems can be traded with $v^2/c^2$, where $v$ is the binary's characteristic velocity and $c$ the speed of light).
%\end{minipage}
%\hfill
%\hskip20pt
%\begin{minipage}[r]{0.68\textwidth}
\begin{figure*}
\centering
\includegraphics[width=0.75\textwidth]{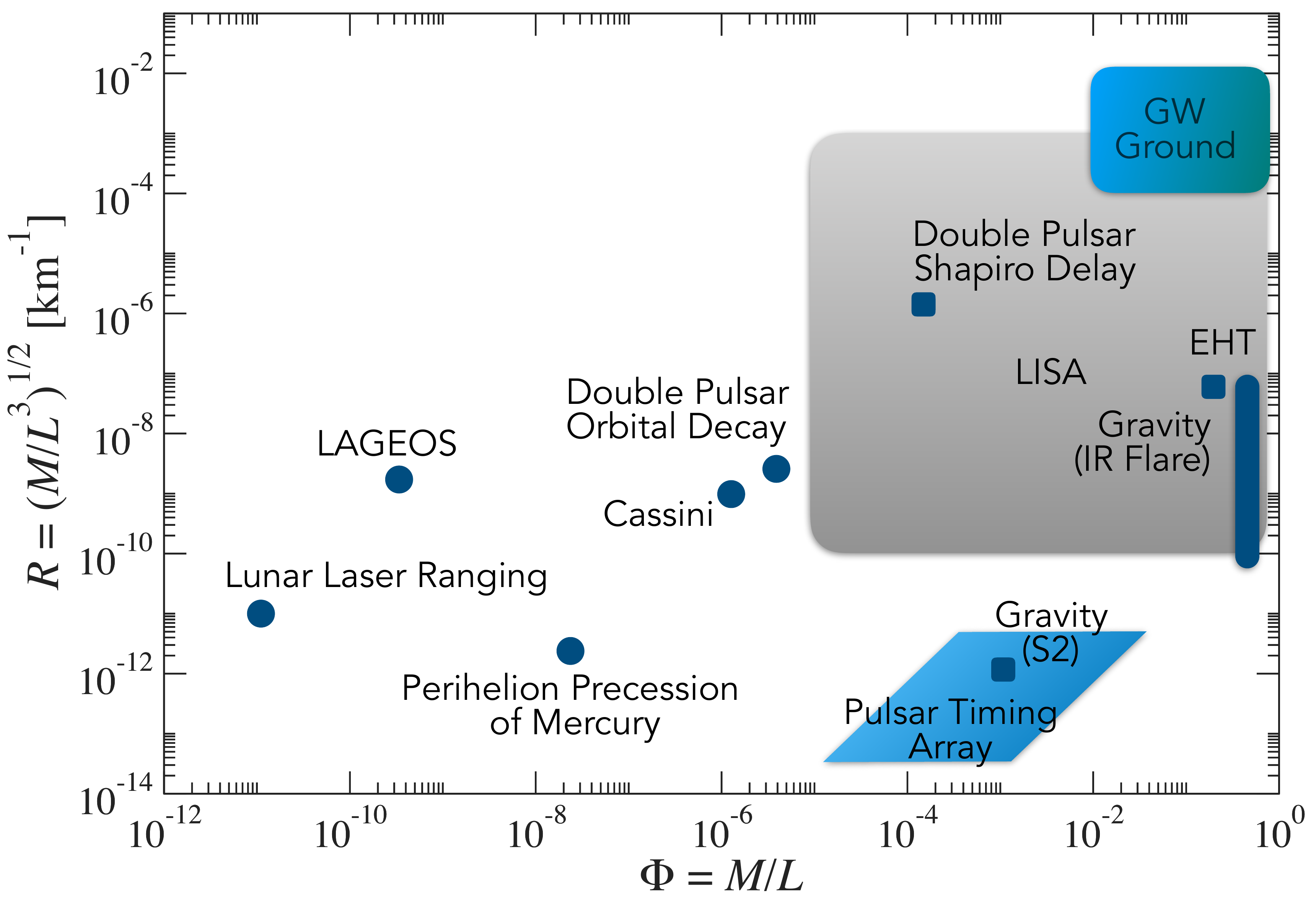}
\caption{ {\bf Probing gravity at all scales:} Illustration of the reach in curvature scales vs potential
scales targeted by different, representative, past/current/future missions. In this figure, $M$ and $L$ are the
characteristic mass and length involved in the observable associated to each mission. For instance, in 
observables associated to binary systems $M$ is the total mass and $L$ the binary's separation, in
this case $M/L$ is related to  $v^2/c^2$ through the virial theorem.}
\label{fig:gravitytests} 
\end{figure*}
%\end{minipage}

\vskip10pt
\noindent{\bf New fields, particles and polarizations} Lovelock’s uniqueness theorem in 4-dimensions ~\cite{Lovelock:1971yv} implies that departures from GR that preserve locality necessarily require the presence of extra degrees of freedom, which generically also arise from theories of quantum gravity in the low-energy limit. This often leads to violations of the strong equivalence principle through the fields’ nonminimal coupling with matter. Among possible theories, those with an additional scalar field are relatively simple~\cite{Brans:1961sx,Fujii:2003pa} yet could give rise to exciting new strong-field phenomenology~\cite{Palenzuela:2013hsa,Shibata:2013pra}. Together with examples of strong-field GW signatures in more complicated scenarios inspired by the low-energy limit of quantum gravity theories~\cite{Okounkova:2017yby,Witek:2018dmd} they also serve as excellent proxies of the type of new physics we can hope to detect. In addition, if a binary’s constituents can become “dressed" with a scalar configuration~\cite{Damour:1993hw, Kanti:1995vq,Mignemi:1992nt, Antoniou:2017acq}, the system emits scalar waves in addition to tensorial ones, with the dominant component being dipolar emission~\cite{Will:2014kxa} (although this may be suppressed for massive fields~\cite{Alsing:2011er,Sagunski:2017nzb}). Extra polarizations can be detected directly~\cite{TheLIGOScientific:2016src}, and indirectly inferred from their effects on the system’s dynamics and consequent impact on GWs~\cite{Will:2014kxa}. 

\noindent{\bf Graviton mass} Recently, the possibility that gravitons could have a mass has resurfaced in theoretical physics within extensions of GR~\cite{deRham:2010kj,Hassan:2011hr}. The current best bound on the graviton mass from LIGO through modified dispersion relations is $m_g < 7.7 \times 10^{-23}\,{\rm eV}/c^2$ ~\cite{TheLIGOScientific:2016src,Abbott:2017vtc} and improvements of two orders of magnitude would be possible with 3G detectors.

\noindent{\bf Lorentz violations} Lorentz symmetry is regarded as a fundamental property of the Standard Model of particle physics, tested to spectacular accuracy in particle experiments~\cite{Mattingly:2005re}. In the gravitational sector, constraints are far less refined. Theories with Lorentz invariance violation (e.g., Horava-Lifschitz~\cite{Horava:2009uw} and Einstein-Aether~\cite{Jacobson:2000xp}) give rise to significant effects on black holes~\cite{Eling:2006ec, Barausse:2011pu}, additional polarizations~\cite{Sotiriou:2017obf}, and the propagation of GWs (e.g. through dispersion and birefringence~\cite{Kostelecky:2016kfm}) which can be greatly constrained by 3G detectors that will observe sources at high redshifts of $z \sim 10$-$20$.

\noindent{\bf Parity violations} Parity violations in gravity arise naturally within some flavors of string theory~\cite{Green:1987mn}, loop quantum gravity~\cite{Ashtekar:1988sw} and inflationary models~\cite{Weinberg:2008hq}. The associated phenomenologies are, to some degree, understood from effective theories~\cite{Jackiw:2003pm}. For instance, they give rise to black holes with nontrivial pseudo-scalar configurations that violate spatial parity~\cite{Yunes:2009hc}. The resulting scalar dipole leads to a correction to the GWs produced through a of a binary inspiral and merger~\cite{Sopuerta:2009iy,Yagi:2012vf,Okounkova:2017yby}. Additionally, parity violating theories can exhibit birefringence, thus impacting the characteristics of GWs tied to their handedness~\cite{Yagi:2017zhb}.

\section*{The nature of dark matter.}
The exquisite ability of 3G detectors to probe the population and dynamics of electromagnetically dark objects  throughout the Universe and harness deep insights on gravity can help reveal the nature of dark matter and answer key questions about its origin.

\noindent{\bf Black holes as dark matter candidates} LIGO and Virgo discoveries have revived interest in the possibility that dark matter could be composed, in part, of black holes of masses $\sim 0.1$–$100\,M_\odot$ ~\cite{Clesse:2016vqa,Bird:2016dcv,Sasaki:2016jop}. Such black holes might have been produced from the collapse of large primordial density fluctuations in the very early Universe or during inflation~\cite{Carr:1974nx,Sasaki:2018dmp}. The exact distribution of masses depends on the model of inflation, and might be further affected by processes in the early Universe such as the quantum-chromodynamic phase transition~\cite{Byrnes:2018clq}.

The detection of GWs from binary systems composed of objects much lighter than stellar mass black holes, or with a mass distribution demonstrating an excess within a certain range, could point towards the existence of primordial black holes~\cite{Abbott:2018oah}. The detection of very high redshift sources would be another hint towards this formation channel~\cite{Koushiappas:2017kqm}. With a sensitivity to observe stellar mass black holes at redshifts of $\sim 10$-$20,$ 3G detectors will be uniquely positioned to determine their mass and spatial distribution, which will be crucial to test this hypothesis~\cite{Kovetz:2016kpi}.

\noindent{\bf Detection of dark matter with compact objects} Beyond probing whether dark matter can be partially made up of black holes, GWs can also scrutinize models where dark matter consists of particles beyond the standard model (e.g., weakly interacting massive particles~\cite{Steigman:1984ac}, fuzzy dark matter~\cite{Hui:2016ltb} or axion-like particles~\cite{Essig:2013lka}). Indeed, binary black holes evolving in a dark-matter rich environment will not only accrete the surrounding material, but also exert a gravitational drag on the dark matter medium, which affects the inspiral dynamics~\cite{Eda:2013gg,Macedo:2013qea,Barausse:2014tra}. Even though their magnitude is small, drag and accretion could have a cumulative effect over a large number of orbits that could be detected by a combination of observatories in space and 3G detectors~\cite{Barack:2018yly}.

Additionally, dark matter that interacts with standard model particles can scatter, lose energy, and be captured in astrophysical objects~\cite{Press:1985ug,Gould:1989gw,Goldman:1989nd,Bertone:2007ae}. The dark-matter material eventually thermalizes with the star, and accumulates inside a finite-size core. The presence of this core might imprint a GW signature on the matter effects during the inspiral and merger of such objects in a binary system~\cite{Ellis:2017jgp}. In certain models, asymmetric dark matter can accumulate and collapse to a black hole in the dense interiors of neutron stars. The core can grow by accumulating the remaining neutron star material, in effect turning neutron stars into light black holes in regions of high dark-matter density such as galactic centers~\cite{Bramante:2017ulk,Kouvaris:2018wnh}. This provides a mechanism for creating light black holes that could be observed by 3G detectors.

\section*{The nature of compact objects.}
Observational evidence so far suggests that compact massive objects in the Universe exist in the form of black holes and neutron stars. Binary systems composed of such objects provide ideal scenarios to unravel both astrophysical and fundamental physics puzzles such as elucidating the connections of strong gravity with the most energetic phenomena in our Universe, exploring the “final state” conjecture~\cite{1969NCimR...1..252P} (i.e., that the end point of gravitational collapse is a Kerr black hole), and probing the existence of horizons.

\noindent{\bf Nature of black holes} Black holes in isolation are the simplest objects in the Universe. Astrophysical black holes are electrically neutral and are described by just two parameters — their mass and spin angular momentum. A perturbed black hole returns to its equilibrium state by oscillating with its characteristic quasi-normal modes, whose frequency and decay time are uniquely determined by the two parameters. By detecting several quasi-normal modes 3G detectors can facilitate multiple null-hypothesis tests of the Kerr metric~\cite{Dreyer:2003bv,Berti:2016lat,Berti:2018vdi}.

\vskip10pt
\hskip-17pt
\begin{minipage}{0.450\textwidth}
{\bf Nature of neutron stars} General relativity, with input from nuclear physics, can describe the structure of ultra-dense neutron stars. However, the neutron star equation of state is currently poorly known \cite{Baym:2017whm}. Knowledge of the equation of state at supranuclear densities facilitated by 3G detectors will provide unprecedented insights on the properties of matter and fundamental interactions in regimes not accessible to laboratory experiments.

\setlength{\parindent}{0.5cm}
Signatures of matter in GWs from a binary inspiral result from a number of effects such as rotational deformations~\cite{Poisson:1997ha}, various kinds of tidal effects including the excitation of internal oscillation modes of the star~\cite{Kokkotas:1995xe,Lai:1993di,Shibata:1993qc,Flanagan:2006sb,Flanagan:2007ix} and spin-tidal couplings~\cite{Landry:2018bil,Abdelsalhin:2018reg}, and the presence of a surface instead of an event horizon~\cite{Hartle:1973zz,Alvi:2001mx,Maselli:2017cmm}. 
\end{minipage}
\hfill
\begin{minipage}{0.50\textwidth}
\includegraphics[width=0.95\textwidth]{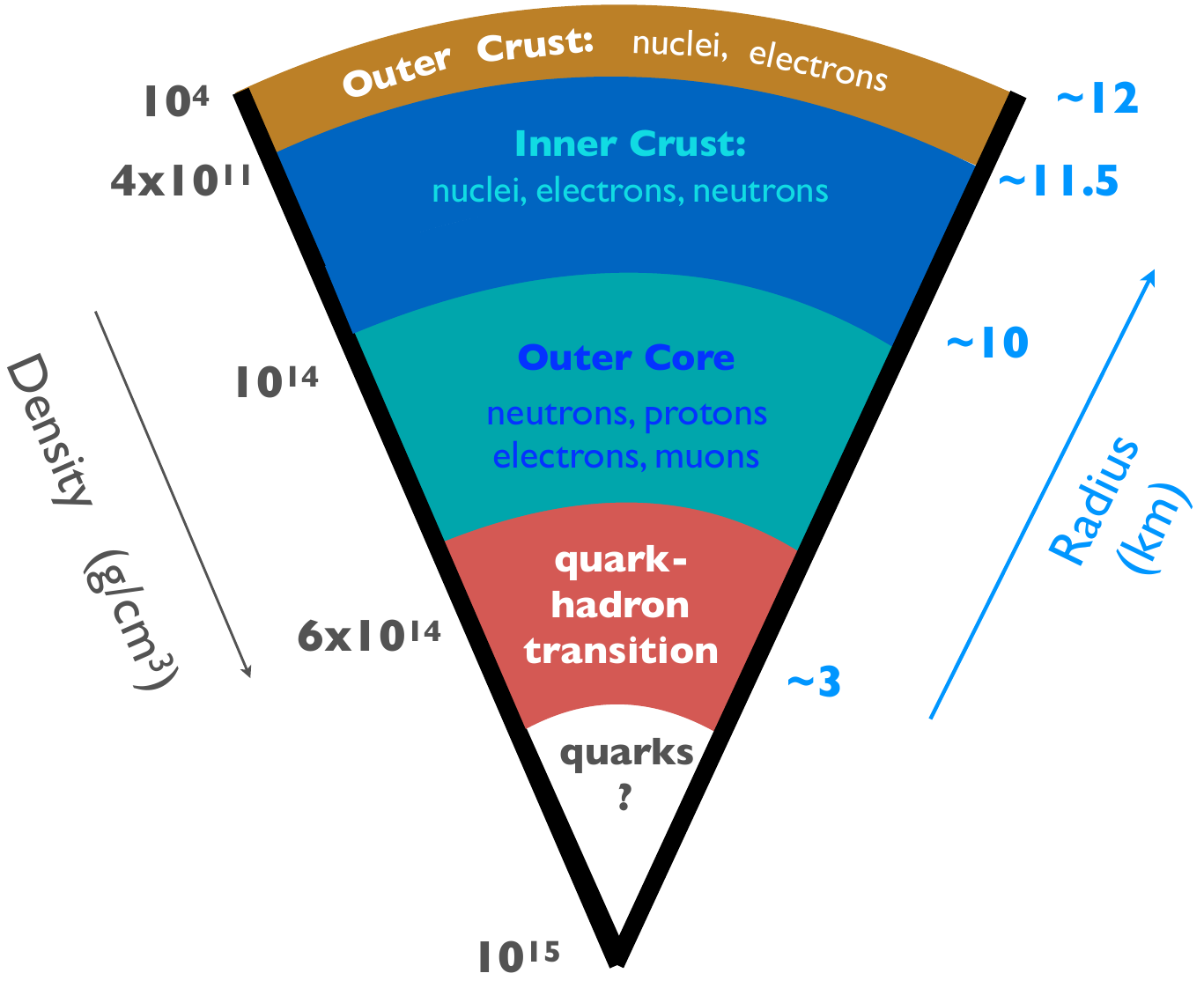}
{\small Internal structure of a neutron star -  predicted by theory. Phase transitions to states 
of matter containing de-confined quarks, hyperons and meson condensates are possible at the densities encountered in the inner core.}
\label{fig:neutronstar_profile}
\end{minipage}

The most striking matter imprints in the waveform occur during the tidal disruption in a neutron star-black hole binary~\cite{Lattimer:1974slx,Shibata:2011jka}, or the merger and post-merger epochs in binary neutron star collisions~\cite{Baiotti:2016qnr}. Signals from these regimes have high frequencies and are therefore very difficult to measure with current detectors. 3G detectors will improve current measurements of tidal deformability by a factor of $\sim$ 10 and thus determine the cold equation of state significantly better, and enable unprecedented measurements of the new physics encountered during the coalescence and post-merger epochs.

\noindent{\bf Beyond black holes and neutron stars} From a phenomenological standpoint, black holes and neutron stars are just two “species” of a larger family of compact objects. More exotic species are theoretically predicted in extensions to GR, but also in particular scenarios within GR~\cite{Cardoso:2017cqb,Barack:2018yly}. For instance, exotic objects arise from beyond-standard model fundamental fields minimally coupled to gravity (e.g., boson stars ~\cite{Liebling:2012fv}), in Grand Unified Theories in the early Universe (e.g., cosmic strings~\cite{Jeannerot:2003qv}), from exotic states of matter, as “dressed” compact objects with further structure stemming from quantum gravitational origin~\cite{Giddings:2013kcj,Giddings:2017mym} or new physics at the horizon scale (e.g., firewalls~\cite{Almheiri:2012rt}), or as horizonless compact objects in a variety of scenarios (e.g., fuzzballs, gravastars, and dark stars~\cite{Mathur:2005zp,Mazur:2004fk, Barcelo:2007yk,Carballo-Rubio:2017tlh,Danielsson:2017riq,Berthiere:2017tms}) .

GW observations provide a unique discovery opportunity in this context, since exotic matter/dark matter might not interact electromagnetically or any electromagnetic signal from the surface of the compact object might be highly redshifted~\cite{Cardoso:2017cqb}. Example GW signatures from the inspiral epoch include dipole radiation as well as the variety of matter effects discussed above in the context of neutron stars~\cite{Barack:2018yly}.

Additionally, while the ringdown signal can be qualitatively similar to that of a black hole, quasi-normal modes of, e.g. gravastars, axion stars and boson stars, are different from Kerr black holes~\cite{Berti:2018vdi}. 3G detectors will have unprecedented ability to extract such modes. In addition to gravitational modes, matter modes might be excited in the ringdown of an extremely compact object, akin to fluid modes excited in a remnant neutron star~\cite{Barack:2018yly}. In the case of certain black hole mimickers the prompt ringdown signal is identical to that of a black hole; however, these objects generically support quasi-bound trapped modes which produce a modulated train of pulses at late time. These modes appear after a delay time whose characteristics are key to test Planckian corrections at the horizon scale that could be explored with 3G detectors~\cite{Cardoso:2017cqb}.

\noindent{\bf Bosonic clouds} Ultralight bosons have been proposed in various extensions of the Standard Model~\cite{Essig:2013lka}. When the Compton wavelength of such light bosons (masses of $10^{-21}$-$10^{-11}$\,eV) is comparable to the horizon size of a stellar or supermassive rotating black hole, superradiance can cause the spin to decay, populating bound Bohr orbits around the black hole with an exponentially large number of particles~\cite{Arvanitaki:2009fg,Pani:2012vp,Brito:2013wya}. Such bound states, in effect “gravitational atoms", have bosonic “clouds" with masses up to $\sim$ 10\% of the mass of the black hole~\cite{Arvanitaki:2010sy,Brito:2014wla,East:2017ovw}. Once formed, the clouds annihilate over a longer timescale through the emission of coherent, nearly-monochromatic, GWs~\cite{Arvanitaki:2010sy,Arvanitaki:2014wva}. 

Alternatively, measuring the spin and mass distribution of binary black holes can provide evidence for characteristic spin down from superradiance~\cite{Arvanitaki:2016qwi,Baryakhtar:2017ngi,Brito:2015oca}, and explore the parameter space for ultralight bosons with 3G detectors. In addition, the presence of such clouds can be probed through the imprint of finite-size
effects on the compact objects in a binary system~\cite{Baumann:2018vus}. GWs will, therefore, provide a unique window into the ultralight, weakly coupled regime of particle physics that cannot be easily probed with terrestrial experiments.

\chapterimage{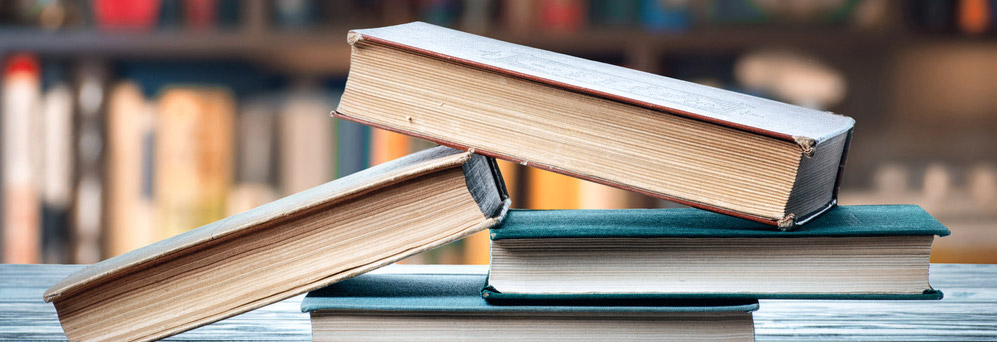} % Chapter heading image
\bibliographystyle{utphys}
\bibliography{wp-all,wp,3g}

\end{document}